\documentclass[fleqn,usenatbib]{mnras}

\usepackage{newtxtext,newtxmath}

\usepackage[T1]{fontenc}
\usepackage{comment}

\DeclareRobustCommand{\VAN}[3]{#2}
\let\VANthebibliography\thebibliography
\def\thebibliography{\DeclareRobustCommand{\VAN}[3]{##3}\VANthebibliography}

\usepackage{graphicx}	
\usepackage{amsmath}	

\newcommand{\Msun}{M$_\odot$}

\title[Simultaneous light and polarisation variability of WR\,40]{
Constraints on Clumps in the Representative Wind of the WN8 Wolf-Rayet star
HD\,96548 = WR\,40 with Simultaneous Broadband Light and Linear-Polarisation Variability}

\author[R. Ignace, A.F.J. Moffat, C. Robert, L. Drissen]{
R. Ignace,$^{1}$\thanks{E-mail: ignace@mail.etsu.edu}
A.F.J. Moffat,$^{2,3}$
C. Robert,$^{3,4}$
and L. Drissen$^{3,4}$
\\
$^{1}$Department of Physics \& Astronomy, East Tennessee State University, Johnson City, TN 37614, USA\\
$^{2}$D\'epartement de physique, Université de Montr\'eal, C.P. 6128, Succ. C-V, Montr\'eal, QC, H3C 3J7, Canada\\
$^{3}$Centre de Recherche en Astrophysique du Qu\'ebec, QC, Canada\\
$^{4}$D\'epartement de physique, de g\'enie physique et d’optique, Universit\'e Laval, Qu\'ebec, QC, G1V 0A6, Canada
}

\date{Accepted XXX. Received YYY; in original form ZZZ}

\pubyear{2022}

\begin{document}
\label{firstpage}
\pagerange{\pageref{firstpage}--\pageref{lastpage}}
\maketitle

\begin{abstract}
We report precision ground-based broadband optical intensity and
linear-polarisation light-curves for the sky's brightest WN8 star,
WR\,40. WN8 stars are notorious for their high level of variability,
stemming from stochastic clumps in their strong winds that are
slower and less hot than the winds of most other Wolf-Rayet (WR)
stars. We confirm previous results that many WR stars display an
amplitude of variability that is an order-of-magnitude higher in
photometry than in polarimetry. For the first time, the unique
nature of near simultaneity of our photometric and polarimetric
observations of WR\,40 allows us to check whether the two types of
variability show correlated behaviour, of which we find none.
Assuming simple temporal functions for the brightness and polarisation
of individual clumps, a model for simulated light curves is found
to reproduce the properties of the observations, specifically the
absence of correlation between photometric and polarimetric variations,
the ratio of standard deviations for photometric and polarimetric
variability, and the ratio of the average intrinsic polarisation
relative to its standard deviation.  Mapping the solution for time
variability to a spatial coordinate suggests that the wind clumps
of WR\,40 are free-free emitting in addition to light scattering.

\end{abstract}





\section{Introduction}

Wolf-Rayet (WR) stars display strong, fast and hot ionized winds arising from spectral line-driving \citep{1993ApJ...405..738L, 1994A&A...289..505S, 1995ApJ...442..296G, 2002A&A...389..162N, 2020MNRAS.499..873S, 2021arXiv210908164V}.  The WR spectral classification is associated with three main types of stars \citep{2007ARA&A..45..177C, 2021ApJS..257...58D}:   (1) main-sequence H-burning stars of very high-mass, above $\sim$60\,\Msun; (2) stars of initial mass $\sim$20-60\,\Msun\ that have evolved to the He-burning stage; and (3) $\sim$15\% of the central stars in planetary nebulae
with initial masses $\sim$1-10\,\Msun. Stars of category (1) are essentially all WNh types. Those of category (2) have subclassifications of WN2-9, WC4-9 or WO1-4, all with hydrogren severely depleted or completely absent. For category (3), the stars are mostly [WC/O], strongly dominating in number over [WN]. Among the categories, the binary fraction is high among (1), medium ($\sim$40\%) among (2) and possibly medium among (3).

Here we concentrate on one well-known star among the second class, HD\,96548 = WR\,40, with type WN8h\footnote{WN8 stars are unusual in their properties, and despite some showing evidence for hydrogen, likely belong to category (2) instead of (1).}. It is the brightest in the sky of the WN8 class, which is well known for being the most intrinsically variable among WR types \citep{Ant95, Mar98}.  Most WN8 stars appear to be single and runaway, likely the result of a SN explosion in a close binary, or slung by close gravitational encounter from a tight young cluster of massive stars 
\citep{Mof89}.

Its relatively high apparent brightness and high level of variability have made WR\,40 a frequent target for detailed study.  The variability of WR\,40 typically at $\sim$0.06 mag rms in the optical and with no evidence of periodicity, is understood to be stochastic in nature, originating in its wind clumping \citep[e.g.][]{Ram19}. Contrary to intuition, the intrinsic stochastic photometric variability of WR stars appears to be highest for the slowest winds \citep[or latest subtypes:][]{Len22}. 
The same applies for spectral line variability \citep{Che20}  
and broadband polarimetric variability \citep{Rob89}. 
The reason for this trend may be attached to a deeper subsurface Fe convection zone at $\sim$170~kK in the cooler WR stars, thus providing more mass and hence larger perturbing gravity waves reaching the stellar hydrostatic surface to generate stochastic wind-clumps \citep{Mic14}. 

\begin{table}	\centering
\caption{WR\,40 basic parameters}%
\label{tab:WR40info}
\begin{tabular}{lc } 
		\hline
		Parameter & Value  \\
		\hline
		RA (J2000) & 11:06:17.20   \\
		DEC (J2000) &$-$65:30:35.24 \\
		Spectral Type & WN8h  \\
		V (mag)  & 7.70  \\
	    B$-$V (mag) & 0.10  \\
	    $D$ (kpc) & 2.80 $\pm$ 0.13  \\
	    $PM$-RA (mas yr$^{-1}$)  & $-$2.196 $\pm$ 0.018 \\
	    $PM$-DEC (mas yr$^{-1}$) & $-$1.745 $\pm$ 0.018 \\
	    Log($L_\ast/L_\odot$) & 5.91 $\pm$ 0.15 \\
	    $M_\ast$ ($M_\odot$) & 28 \\  
	    $R_\ast$ ($R_\odot$) & 14.5 $\pm$ 1.5 \\
	    $T_\ast$ (kK) & 44.7 $\pm$ 3.0 \\
	    v$_\infty$ (km s$^{-1}$)$^\dag$ & 840 $\pm$ 130 \\
	    $\dot{M}$ ($M_\odot$ yr$^{-1}$) & (6.3 $\pm$ 1.9) $\times$ 10$^{-5}$ \\
		\hline
	\end{tabular}
	
	{\small $^\dag$ The terminal speed is taken from \citet{Her01} and mistakenly reported as 650 km s$^{-1}$ in \cite{Ham19}.}
\end{table}

\begin{figure*}
	\includegraphics[width=1.5\columnwidth]{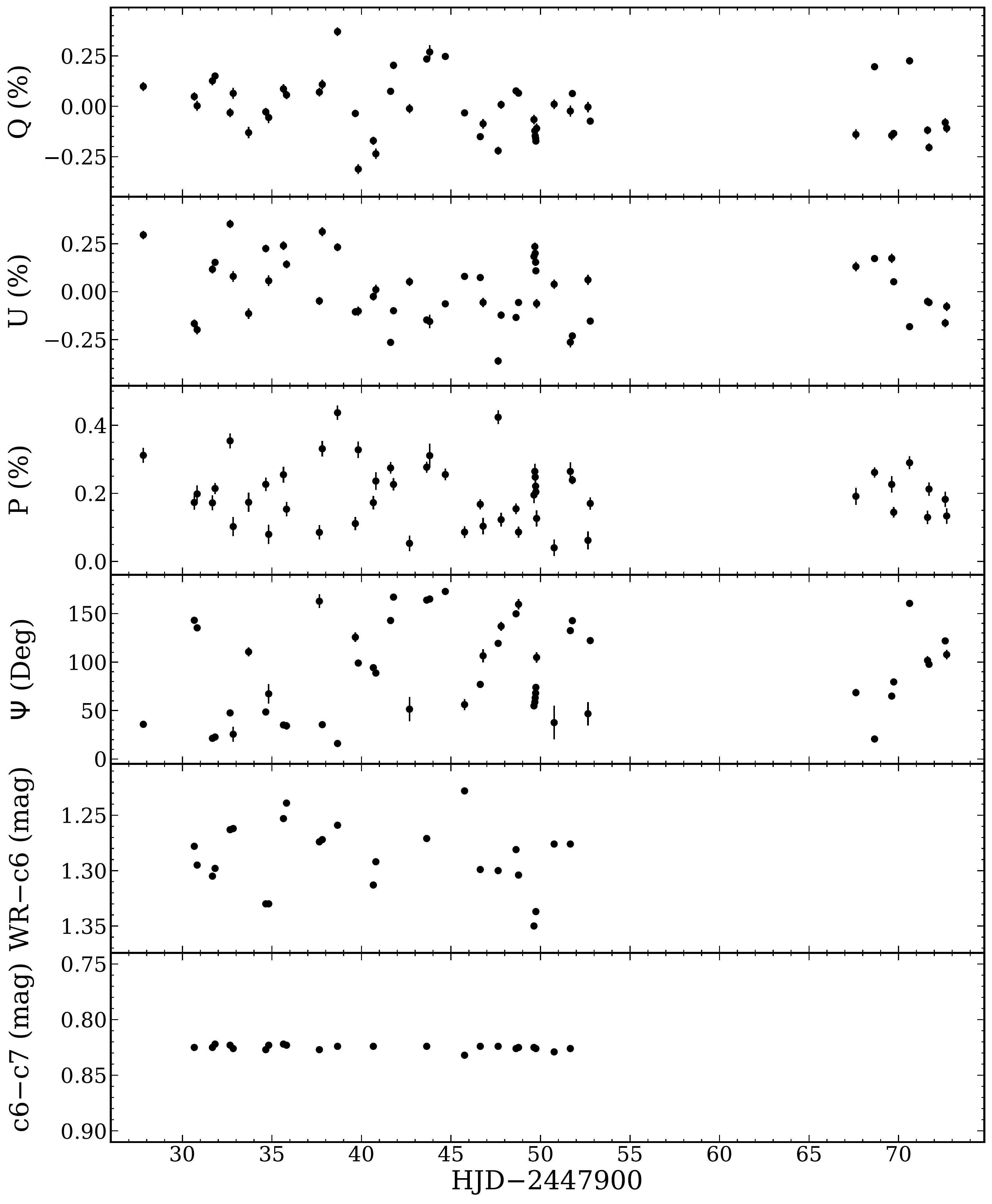}
    \caption{Intrinsic polarisation and differential magnitudes}
    \label{fig:netdata}
\end{figure*}

The paper is structured as follows.  Section~\ref{sec:obs} describes our observations of WR\,40 obtained in photometry and linear polarimetry, including how the data were corrected for interstellar polarisation.
A numerical model for simulating both photometric and polarimetric light curves is presented in Section~\ref{sec:model}.  In this section model parameters are set by the requirement of reproducing the statistical characteristics of the observed variability.  

\begin{table*}	\centering
\caption{Observations}%
\label{tab:WR40obs}
\begin{tabular}{lccccccc} 
		\hline
		HJD	&Q$_{\rm obs}$ 	&U$_{\rm obs}$ 	&P$_{\rm obs}$ 	&$\sigma$ &$\Psi_{\rm obs}$	&c6$-$c7 &WR$-$c6  \\
		$-$2447900	&\%	&\% &\%	&\%		&Deg	&mag	&mag \\
		\hline
27.793 &$-$0.615 &$-$0.718 &0.945  &0.022 &114.7 &--- &---    \\
30.640 &$-$0.665 &$-$1.180 &1.354  &0.021 &120.3 &0.825 &1.278 \\
30.800 &$-$0.711 &$-$1.212 &1.405  &0.025 &119.8 &--- &1.295   \\
31.654 &$-$0.587 &$-$0.897 &1.072  &0.022 &118.4 &0.825 &1.305 \\
31.800 &$-$0.563 &$-$0.861 &1.029  &0.016 &118.4 &0.822 &1.298 \\
32.640 &$-$0.745 &$-$0.661 &0.996  &0.022 &110.8 &0.823 &1.263 \\
32.814 &$-$0.649 &$-$0.934 &1.138  &0.028 &117.6 &0.826 &1.262 \\
33.675 &$-$0.844 &$-$1.128 &1.409  &0.028 &116.6 &--- &--- \\
34.633 &$-$0.741 &$-$0.789 &1.083  &0.020 &113.4 &0.827 &1.330 \\
34.796 &$-$0.769 &$-$0.957 &1.228  &0.028 &115.6 &0.823 &1.330 \\
35.620 &$-$0.627 &$-$0.774 &0.996  &0.023 &115.5 &0.822 &1.253 \\
35.792 &$-$0.657 &$-$0.871 &1.091  &0.021 &116.5 &0.823 &1.239 \\
37.626 &$-$0.643 &$-$1.062 &1.242  &0.021 &119.4 &0.827 &1.274 \\
37.789 &$-$0.605 &$-$0.701 &0.926  &0.023 &114.6 &--- &1.272 \\
38.642 &$-$0.343 &$-$0.782 &0.854  &0.021 &123.0 &0.824 &1.259 \\
39.636 &$-$0.749 &$-$1.119 &1.346  &0.019 &118.1 &--- &--- \\
39.800 &$-$1.025 &$-$1.115 &1.515  &0.024 &113.7 &--- &--- \\
40.642 &$-$0.884 &$-$1.039 &1.364  &0.020 &114.8 &0.824 &1.313 \\
40.782 &$-$0.949 &$-$1.003 &1.381  &0.026 &113.3 &--- &1.292 \\
41.605 &$-$0.639 &$-$1.275 &1.426  &0.017 &121.7 &--- &--- \\
41.768 &$-$0.510 &$-$1.113 &1.224  &0.018 &122.7 &--- &--- \\
42.661 &$-$0.725 &$-$0.962 &1.204  &0.023 &116.5 &--- &--- \\
43.622 &$-$0.479 &$-$1.161 &1.256  &0.016 &123.8 &0.824 &1.271 \\
43.790 &$-$0.444 &$-$1.169 &1.251  &0.035 &124.6 &--- &--- \\
44.662 &$-$0.466 &$-$1.077 &1.173  &0.017 &123.3 &--- &--- \\
45.737 &$-$0.746 &$-$0.934 &1.195  &0.017 &115.7 &0.832 &1.228 \\
46.614 &$-$0.864 &$-$0.940 &1.277  &0.015 &113.7 &0.824 &1.299 \\
46.772 &$-$0.800 &$-$1.070 &1.336  &0.024 &116.6 &--- &--- \\
47.613 &$-$0.934 &$-$1.375 &1.662  &0.020 &117.9 &0.824 &1.300 \\
47.779 &$-$0.705 &$-$1.136 &1.337  &0.020 &119.1 &--- &--- \\
48.613 &$-$0.637 &$-$1.148 &1.313  &0.016 &120.5 &0.826 &1.281 \\
48.752 &$-$0.648 &$-$1.070 &1.251  &0.016 &119.4 &0.825 &1.304 \\
49.615 &$-$0.779 &$-$0.830 &1.138  &0.024 &113.4 &0.825 &1.350 \\
49.663 &$-$0.835 &$-$0.779 &1.142  &0.022 &111.5 &--- &--- \\
49.683 &$-$0.860 &$-$0.814 &1.184  &0.022 &111.7 &--- &--- \\
49.707 &$-$0.872 &$-$0.860 &1.225  &0.019 &112.3 &--- &--- \\
49.721 &$-$0.886 &$-$0.905 &1.267  &0.013 &112.8 &0.826 &1.337 \\
49.762 &$-$0.823 &$-$1.076 &1.355  &0.024 &116.3 &--- &--- \\
50.742 &$-$0.703 &$-$0.975 &1.202  &0.024 &117.1 &0.829 &1.276 \\
51.646 &$-$0.737 &$-$1.277 &1.474  &0.027 &120.0 &0.826 &1.276 \\
51.760 &$-$0.650 &$-$1.244 &1.404  &0.012 &121.2 &--- &--- \\
52.631 &$-$0.717 &$-$0.952 &1.192  &0.026 &116.5 &--- &--- \\
52.759 &$-$0.787 &$-$1.167 &1.387  &0.018 &118.0 &--- &--- \\
67.598 &$-$0.853 &$-$0.883 &1.228  &0.025 &113.0 &--- &--- \\
68.640 &$-$0.517 &$-$0.841 &0.987  &0.015 &119.2 &--- &--- \\
69.598 &$-$0.858 &$-$0.840 &1.201  &0.024 &112.2 &--- &--- \\
69.710 &$-$0.848 &$-$0.962 &1.282  &0.015 &114.3 &--- &--- \\
70.596 &$-$0.488 &$-$1.196 &1.292  &0.019 &123.9 &--- &--- \\
71.600 &$-$0.832 &$-$1.065 &1.352  &0.020 &116.0 &--- &--- \\
71.681 &$-$0.918 &$-$1.070 &1.410  &0.020 &114.7 &--- &--- \\
72.587 &$-$0.794 &$-$1.177 &1.420  &0.022 &118.0 &--- &--- \\
72.668 &$-$0.822 &$-$1.091 &1.366  &0.023 &116.5 &--- &--- \\
		\hline
	\end{tabular}
\end{table*}

\section{Observations}  \label{sec:obs}

Table~\ref{tab:WR40info} provides several basic parameters for WR\,40: position (RA, DEC), 
spectral type, magnitude V, colour-index B$-$V, Gaia EDR3 distance ($D$) and proper motion ($PM$),
stellar luminosity $L_\ast$, mass $M_\ast$, radius $R_\ast$, and hydrostatic-surface temperature $T_\ast$, along with the wind terminal velocity v$_\infty$ and mass-loss rate $\dot{M}$ 
\citep[according to Simbad and][]{Ham19}. 

Broadband linear polarimetry was obtained over an interval of 45 days in February/March 1990 with the Minipol polarimeter of the University of Arizona \citep[see][]{StL87} attached to the 0.6~m University of Toronto telescope on Las Campanas, Chile (now moved to Cerro Burek as part of el Leoncito Observatory, Argentina). The 52 polarimetric observations (Q$_{\rm obs}$, U$_{\rm obs}$, P$_{\rm obs}$, and $\Psi_{\rm obs}$) were obtained on 29 nights (see Tab.~\ref{tab:WR40obs}), spread out in pairs on 18 of those nights and 6 observations on one of the nights. More data from the same observing run were published by \cite{1992ApJ...386..288D} (WR14, WR25 and WR69) and \cite{1992ApJ...397..277R} (WR6).

A standard G-band filter \citep[with central wavelength 4640~\AA\ and bandwidth 1280~\AA; see][]{Bes05} was used for the polarimetric observations of WR\,40. Typical precision of each data-point is $\sigma$(P$_{\rm obs}$) = $\sigma$(Q$_{\rm obs}$) = $\sigma$(U$_{\rm obs}$) = 0.021\%. After the gap between HJD\,2447952 and HJD\,2447967, 0${\buildrel \circ \over .}$4 was subtracted from $\Psi_{\rm obs}$ to be consistent with the data before the gap, based on polarised standard stars (HD\,110984, HD\,111579, HD\,126593 and HD\,147084; the unpolarised standard stars HD\,98161 and 61\,Vir have been observed as well). 

\begin{figure}
	\includegraphics[width=\columnwidth]{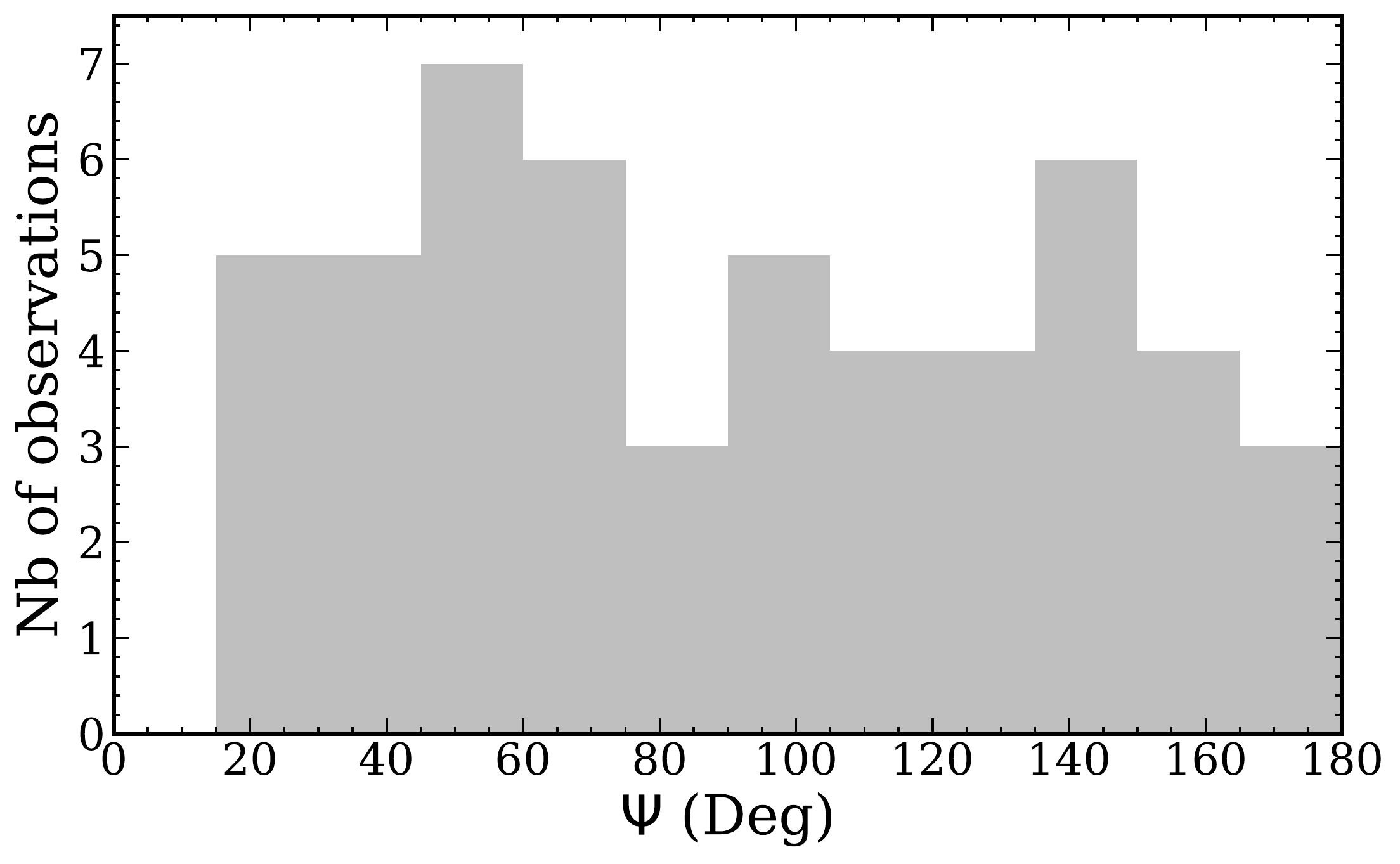}
    \caption{Histogram for polarisation position angle $\Psi$. Note that the observations determine 2$\Psi$ between 0 and 360$^\circ$, hence $\Psi$ from 0 to 180$^\circ$.}
    \label{fig:Thisto}
\end{figure}

Differential photometry was obtained also with Minipol relative to two proven constant comparison stars, HD\,96568 and HD\,96287, noted as c6 and c7 resp.\ by \citet{LaM87}.  Observations were obtained on 16 of the 29 nights with 25 measurements, before the 15-day time gap (see Tab.~\ref{tab:WR40obs}). A 0.89 neutral density filter was added to the G-band filter for the photometry. Based on the differences c6$-$c7, the external error of each photometric data point for WR$-$c6 is 0.003~mag rms. The typical duration of each photometric data-point was 30~minutes, with the sequence sky-c6-WR-c7-WR-c6-sky, following the 20-minute observation used to secure the polarimetric data. This delay between polarimetry and photometry is negligible compared to the timescale of the intrinsic variations, allowing us to classify the observations as essentially simultaneous. 

Figure~\ref{fig:netdata} displays the intrinsic polarimetric parameters Q, U, P, and $\Psi$ and photometric observations WR$-$c6 and c6$-$c7 as functions of time.
The intrinsic polarisation was obtained after subtraction of the interstellar (ISM) polarisation simply given by the weighted average of all the raw Q$_{\rm obs}$ and U$_{\rm obs}$ values (i.e. Q$_{\rm ISM} = -0.713 \pm 0.003$\% and
U$_{\rm ISM} = -1.014 \pm 0.003$\%, or 
P$_{\rm ISM} = 1.239 \pm 0.003$\% and $\Psi_{\rm ISM} = 117.4 \pm 0{\buildrel \circ \over .}1$).  These values for interstellar 
polarisation compare well with those derived by \cite{Dri87} for WR~40 based on averages of several surrounding field stars.

As shown in Figure~\ref{fig:netdata},  Q, U, and P change with time with a maximum peak-to-valley amplitude of $\sim$0.5\% through all possible angles between $\Psi = 0$ and 180$^\circ$ (see Fig.~\ref{fig:Thisto}).  The low value in Figure~\ref{fig:Thisto} for the smallest angles is likely due to small number statistics.
Also, the c6$-$c7 difference magnitude is quite flat (rms = 0.003 mag), confirming the photometric variablity of WR\,40 with a maximum peak-to-valley amplitude $\sim$0.14~mag (and a rms = 0.030~mag).

\begin{figure}
	\includegraphics[width=\columnwidth]{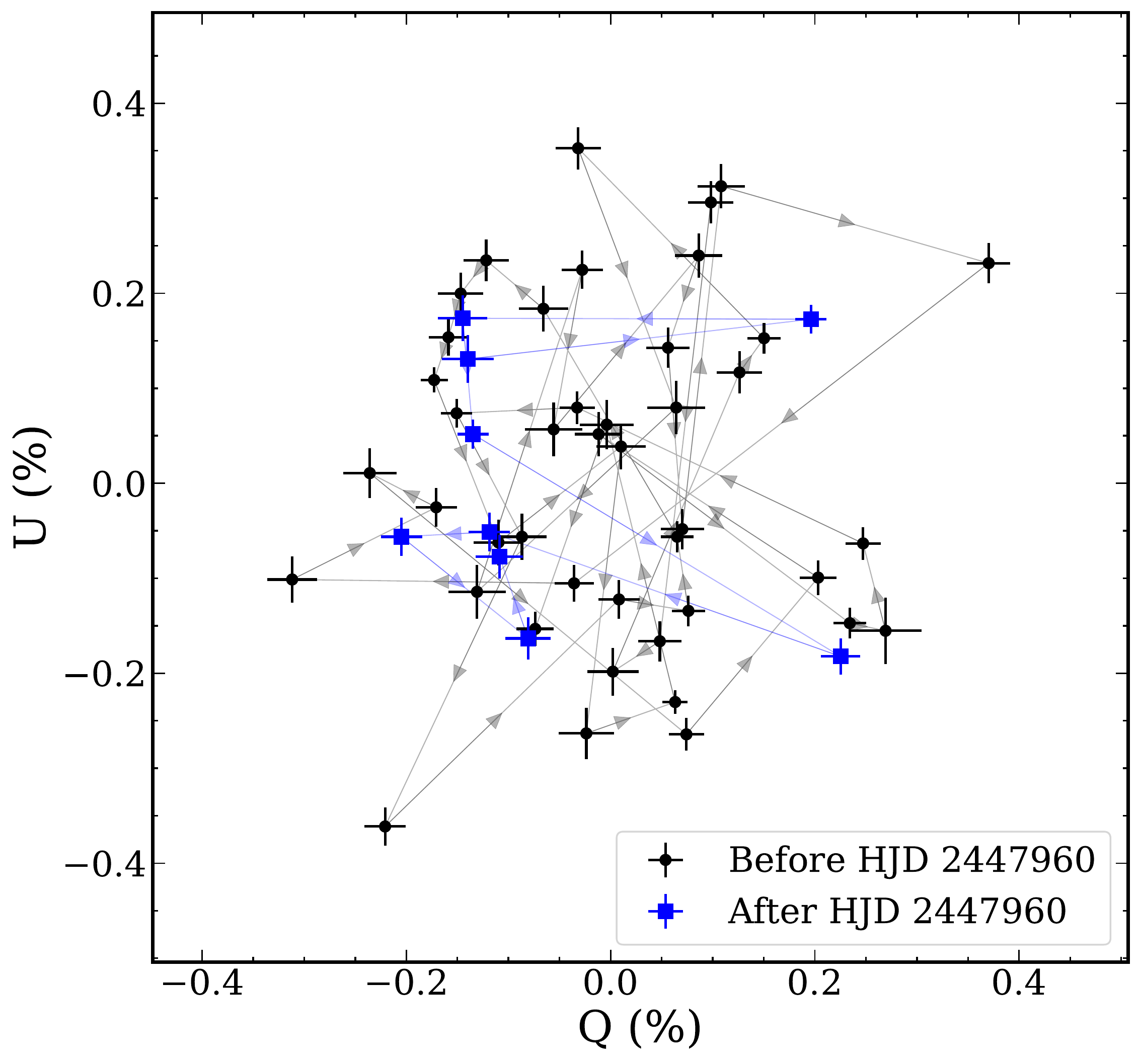}
    \caption{Intrinsic polarisation Q {\it vs} U. The two observing groups are distinguished by different colours. Arrows reflect the observing time-sequence.}
    \label{fig:QUint}
\end{figure}

The random nature of the polarimetric variability is 
also well illustrated by the Q {\it vs} U plot in Figure~\ref{fig:QUint}. Arrows in this plot go from one point to the next following the time sequence of the observations. A periodicity search was carried out among the polarimetric and photometric data separately, with negative results, consistent with the later more numerous precision space photometry for WR\,40 \citep{Ram19}.  The similarity with the plot shown in Figure~12 of \cite{Dri87} is striking, in terms of total amplitude, variability rate and overall shape; these data were obtained with the same equipment as the ones presented here, four years before. Note however the offset caused by the subtraction of the interstellar component in the present paper.

\begin{figure}
	\includegraphics[width=\columnwidth]{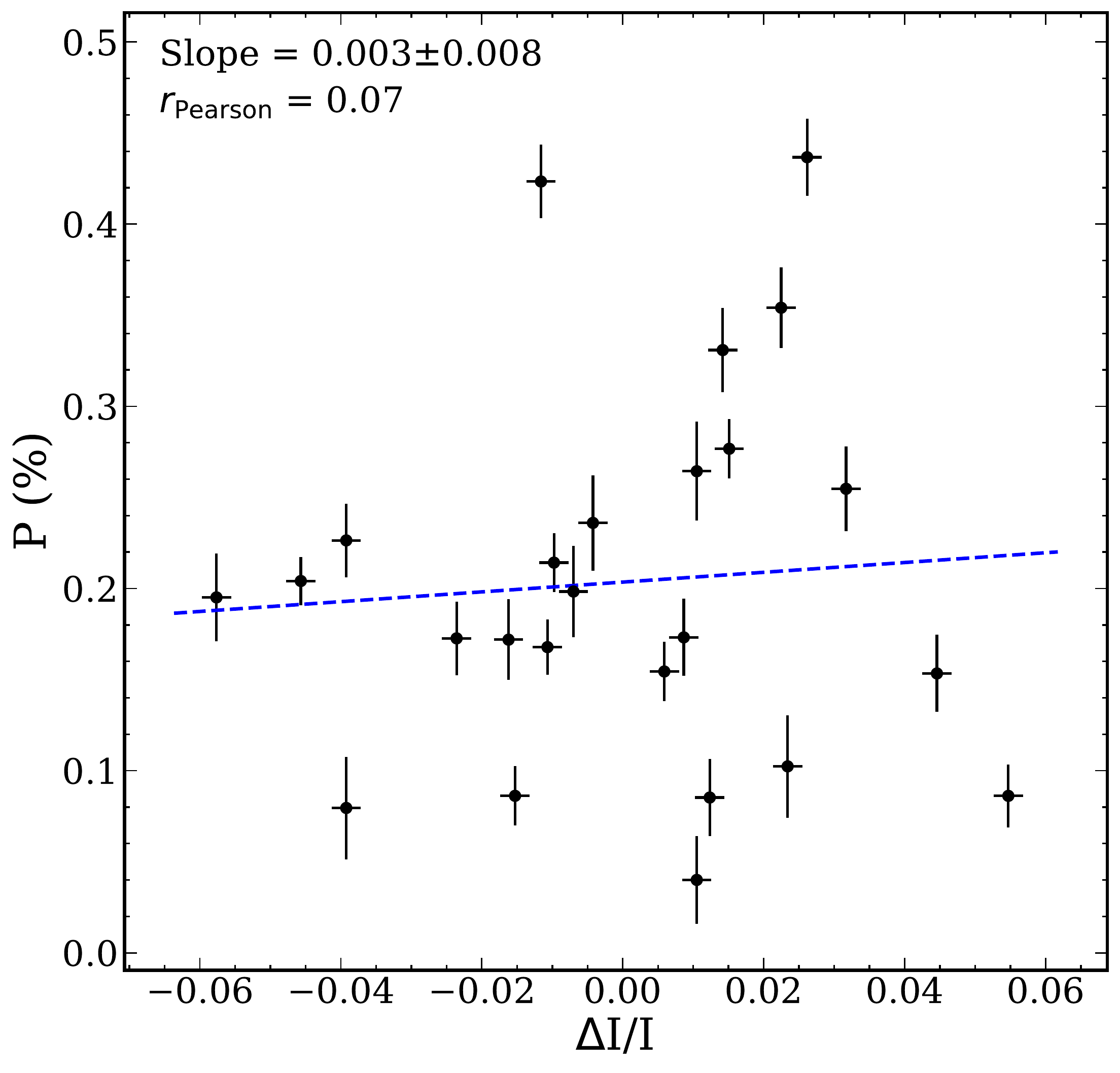}
    \caption{Polarimetric {\it vs} photometric data, for simultaneous observations. The photometric data have been converted to relative intensity: $\Delta$I/I~=~$-$\{(WR$-$c6)$-$$\overline{( {\rm WR}\!-\!{\rm c6})}$\}/1.086. 
    The blue dotted line is a linear regression fit to the data (the corresponding slope value and Pearson correlation coefficient,  $r_{\rm Pearson}$, are indicated at the top of the plot).}
    \label{fig:Pmag}
\end{figure}

The surprisingly uncorrelated behaviour of the polarimetric {\it vs} photometric data is shown in Figure~\ref{fig:Pmag}. These observations give a ratio of the photometric variation (i.e. standard deviation) over the polarimetric variation ${\cal R} = \sigma_{\rm m}/\sigma_{\rm P} \simeq 27$. Naively, one might expect larger changes in P and smaller changes in magnitude at the stellar limbs and the opposite towards the centre of the star. This will be explored in more detail below.

\section{Model for Variable Photometry and Polarimetry}
\label{sec:model}

We explore a model to interpret the data presented in the preceding section.  A model must account for several properties of the observations.  One is that the Q-U plot of Figure~\ref{fig:QUint} is largely a scatter plot, suggestive of two characteristics for the wind structure of WR\,40.  First, the unresolved star is spherically symmetric in time average.  Second, the variability is a stochastic process.  If either of these were invalid, one would expect variable polarisation either at fixed position angle, like Be stars with circumstellar disks, or loop patterns traced out in time, as results for binary star systems \citep[cf., Fig.~3 of][]{2006ASPC..355..173D}.

Other properties involve statistical moments of the light curves.   Variations in photometry, $\Delta {\rm m}$, can be defined to have a zero mean with time, and fluctuations about the mean will exhibit the standard deviation $\sigma_{\rm m}$.  For the polarimetric variations, while average $\overline{\rm Q}=\overline{\rm U}=0$, the average polarisation $\overline{\rm P}$ is nonzero, which must be reproduced by the model.  Additionally, the fluctuations in the polarisation, $\sigma_{\rm P}$, likewise must be reproduced.  The final property of the data is the lack of correlation between photometric and polarimetric light curves, as shown in Figure~\ref{fig:Pmag}. 

\citet{Bro95} explored models to reproduce the mean value, fluctuation, and lack of correlation of WR photometric and polarimetric light curves based on non-simultaneous data in the context of wind clumps that were explicitly optically thin to electron scattering.  Their models were unable to reproduce the observed ratio ${\cal R} \simeq 20$ gathered at the time for WR\,40 using the photometric and polarimetric data of \citet{Dri87} and \citet{LaM87}, respectively at different epochs. This ratio was believed to be typical of WR stars in general based on the work of \citet{Rob92} and \citet{MoR92}. The models of \citet{Bro95} gave lower ratios by a factor of about 4.  Interpreting this shortfall, they offered   explanations for additional physics consider, along with estimates for quantitative impact:  (1) multiple clumps at randomly different positions in the wind, which will reduce the net level of polarisation variability due to geometric cancellation; (2) enhanced recombination emission from clumps due to higher density than outside the clumps in unpolarised light; and (3) multiple scattering within clumps, which will reduce the net polarisation of scattered light off any clump. 

\cite{Ric96} explored in more detail the effects of considering an ensemble of thin-scattering clumps and reduced the factor of disagreement from $4\times$ to about $3\times$.  \cite{2000A&A...357..233L} appear capable of recovering the observed ratio $\cal{R}$ through a consideration of wind velocity law effects not included in \cite{Ric96}; however, those models imply fairly high mass-loss rates of $10^{-4} M_\odot~{\rm yr}^{-1}$.  \cite{2000ApJ...540..412R} considered multiple scattering in thick clumps using Monte Carlo radiative transfer modelling.  While it appears that their models can reproduce $\cal{R}$ ratios, the results for $\overline{\rm P}$ are too large by about $2\times$, plus no synthetic light curves are presented with relevance for the observed timescale of variability.

We consider a different approach to the problem.  In \citet{Ram19}, a strictly temporal function was adopted for the brightness distribution of an individual clump
in WR\,40.  Simulated light curves were generated to model observations made with BRITE.  A match to the statistical characteristics of the measurements, as constrained by the adopted mass-loss rate, led to a brightness amplitude for the clumps, an estimate of clump mass, and inference of a time constant.  We extend the approach of \citet{Ram19} with the inclusion of another temporal function to represent the polarimetric variability, as described next.

\subsection{Description of the Model}

In \citet{Ram19}, clumps were assumed to emerge from a pseudo-photosphere formed in the wind, that was the location of continuum formation at optical wavelengths.  Assuming a smooth wind, the radius, $R_{\rm phot}$, of this pseudo-photosphere was estimated from the condition of optical depth unity in electron scattering, noting that electron scattering is linear in density and not subject to the clumping biases that influence density-squared diagnostics \citep[e.g.,][]{1991A&A...247..455H, 2008cihw.conf.....H}.  A value of $R_{\rm phot} \approx 2.4\,R_\ast$ was obtained.  For the temporal function, we reproduce here Equation~5 from that paper:

\begin{equation}
\frac{f_{\rm s}}{f_\ast} = \left(\frac{f_0}{f_\ast}\right)\,H(\Delta t) \,e^{-\Delta t^2/\tau^2} ,   
\label{eq:fs}
\end{equation}

\noindent where $f_{\rm s}$ is the flux of scattered light by the clump, $f_\ast$ is the photospheric flux that is considered constant, $f_0$ is the brightness of the clump at the photosphere, $\Delta t$ is the travel time in the wind of the clump after emerging from the photosphere, $H$ is the Heaviside function which is zero for $\Delta t < 0$ and unity otherwise, and $\tau$ is a free parameter of the model.  This time profile function is the declining half of a Gaussian, with half-width at half maximum of $\Delta t_{1/2} = 0.83\,\tau$.  The function was introduced to fit light-curve measurements of  WR\,40 obtained with {\em BRITE}.  A value of $\tau \approx$ 21.6 hours was obtained, in the case of all clumps being identical (Case B of that paper; see below).

Another key free parameter for modelling the BRITE data was the injection rate of clumps.  Assuming the wind consists entirely of clumps \citep[e.g.,][]{2020Galax...8...60H}, the injection rate is related to the mass-loss rate, $\dot{M}$, of the wind and used to derive clump masses, with

\begin{equation}
\dot{M} = \overline{m}_{\rm c}\,\dot{\cal N},
\label{eq:mdot}
\end{equation}

\noindent where $\overline{m}_{\rm c}$ is the average clump mass and $\dot{\cal N}$ is the clump injection rate into the wind.  \citet{Ram19} considered two types of models:  one in which all clumps had the same mass, and one that used a turbulent power-law distribution of clump masses, with progressively more clumps of small mass.  Simulated light curves matched the photometric data equally well by both clump models.  

Our model involves clumps that emerge from the pseudo-photosphere formed in the wind.  Upon emerging, the clump has its peak brightness, which then declines monotonically with time, as the clump moves away from the star.  However, for the polarisation the behaviour is different, with a clump having initially zero polarisation.  As the clump moves away from the star, the polarisation rises, peaks, and then declines.

This kind of behavior is expected for a uniformly bright stellar atmosphere. \citet{Cas87} derived the finite star depolarisation factor for thin electron scattering polarisation and showed that polarisation at the photosphere is zero (due to hemispherical isotropy of the illuminating radiation), increases to a peak value at a radius $r_{\rm peak} > R_{\rm phot}$, and thereafter declines with larger radius.  This behaviour is associated with a uniformly bright atmosphere; the effect of non-uniform stellar surface brightness can change the trend \citep[e.g.,][]{1989ApJ...344..341B}.
To replicate qualitatively the case of a uniformly bright photosphere, the following form for the flux amplitude of linearly polarised light $f_{\rm P}$ from a single clump is introduced:

\begin{equation}
\frac{f_{\rm P}}{f_\ast} = \left(\frac{f_L}{f_\ast}\right)\,H(\Delta t)\,K_\alpha^{-1}\,\left(\Delta t / \tau\right)^\alpha \,e^{-\Delta t^2/\tau^2},  
\label{eq:fp}
\end{equation}

\noindent where $f_L$ is the peak level of linearly polarised flux, the exponent $\alpha$ is larger than 0, 
and $K$ is a normalisation constant that depends on the value of $\alpha$.  The required normalisation is given by

\begin{equation}
K_\alpha= \alpha^{\alpha/2}\,e^{-\alpha/2}.
\end{equation}

To obtain Stokes-Q and U fluxes from an individual clump, a trajectory for the clump must be specified in terms of a position angle on the sky $\phi$, and a polar angle $\theta$ from the observer axis.  We obtain fractional Q and U parameters as

\begin{eqnarray}
{\rm Q} & = & \left( \frac{f_{\rm P}} {f_\ast} \right)\,\sin^2\theta \cos 2\phi,~{\rm and} \\
{\rm U} & = & \left( \frac{f_{\rm P}} {f_\ast} \right)\,\sin^2\theta \sin 2\phi.
\end{eqnarray}

We define a model run for a duration of time $T$ with a discrete number of clumps $N_T$.  These selections lead to an average injection rate $\dot{{\cal N}}=N_T/T$.  We assign random times for the appearance of each clump at the photosphere as $t_{\rm j} = s \times T$, where $s \in [0,1]$ is a random number.  The wind is assumed spherically symmetric in time average.  This implies that the azimuth has no preferred direction, hence we also sample $\phi = 2\pi \times s$ randomly.  Random directions in solid angle require 
that the polar angle be sampled according to $\cos \theta = -1 + 2\,s$.  

At time $t$ in the synthetic light curve, the scattered light and polarisation properties of the model are determined from:

\begin{eqnarray}
f_{\rm tot}(t) & = & \sum_{\rm k} \, f_{\rm s}(\Delta t_{\rm k}), \\
{\rm Q}_{\rm tot}(t) & = & \sum_{\rm k} \, {\rm Q}(\Delta t_{\rm k}), \\
{\rm U}_{\rm tot}(t) & = & \sum_{\rm k} \, {\rm U}(\Delta t_{\rm k}), \\
{\rm P}(t) & = & \sqrt{{\rm Q}^2_{\rm tot} + {\rm U}^2_{\rm tot}}, \\ 
\tan 2\Psi(t) & = & \frac{{\rm U}_{\rm tot}}{{\rm Q}_{\rm tot}}.
\end{eqnarray}

\noindent The random sampling over the light-curve duration, $T$, leads to a set of times $\{ t_{\rm k}\}$ at which the ensemble of clumps k\,$=1,\dots,N_T$ appear at the photosphere.  At any time time $t$, the k$^{th}$ clump will have been in the wind for an interval $\Delta t_{\rm k} = t-t_{\rm k}$, for those clumps with $\Delta t_{\rm k} \ge 0$.

We characterise the statistical properties of the simulated light curve with time-averaged values as follows: 

\begin{eqnarray}
\overline{f} & = & \frac{1}{{\rm I}_{\rm max}}\,\sum_{1}^{{\rm I}_{\rm max}}\, f_{\rm tot}(t_{\rm i}), \\
\overline{\rm Q} & = & \frac{1}{{\rm I}_{\rm max}}\,\sum_{1}^{{\rm I}_{\rm max}}\, {\rm Q}_{\rm tot}(t_{\rm i}), \\
\overline{\rm U} & = & \frac{1}{{\rm I}_{\rm max}}\,\sum_{1}^{{\rm I}_{\rm max}}\, {\rm U}_{\rm tot}(t_{\rm i}), \\
\overline{\rm P} & = & \frac{1}{{\rm I}_{\rm max}}\,\sum_{1}^{{\rm I}_{\rm max}}\, {\rm P}_{\rm tot}(t_{\rm i}),
\end{eqnarray}

\noindent where the values $\{t_{\rm i}\}$ represent the discrete time samples of the model light-curve, of which there are ${\rm I}_{\rm max}$ measures in total.  To be clear, the set \{k\} is an index for the clumps and $t_{\rm k}$ refers to the times when clumps appear at the $R_{\rm phot}$ (and so are observable); whereas the set \{i\} refers to the synthetic observations and $t_{\rm i}$ refers to the sampling of the model light curve. The expectation is that for spherical symmetry, both $\overline{\rm Q}$ and $\overline{\rm U}$ will tend toward zero; however, $\overline{\rm P}$ will generally be nonzero \citep[e.g.,][]{Bro95}.  Additionally, the average scattered light $\overline{f}$ is also nonzero, which serves as a bias offset to the assumed non-varying photospheric flux $f_\ast$.

Standard deviations associated with light fluctuations for photometry, or flux, and the polarimetry are obtained from the observational dataset and provide constraints for the model.  For the synthetic light curves, we introduce standard deviations for the model data as:

\begin{eqnarray}
\sigma_{f}^2 & = & \frac{1}{{\rm I}_{\rm max}}\,\sum_{1}^{{\rm I}_{\rm max}}\, \left[f_{\rm tot}(t_{\rm i})-\overline{f}\right]^2, \\
\sigma_{\rm Q}^2 & = & \frac{1}{{\rm I}_{\rm max}}\,\sum_{1}^{{\rm I}_{\rm max}}\, \left[{\rm Q}_{\rm tot}(t_{\rm i})-\overline{\rm Q}\right]^2, \\
\sigma_{\rm U}^2 & = & \frac{1}{{\rm I}_{\rm max}}\,\sum_{1}^{{\rm I}_{\rm max}}\, \left[{\rm U}_{\rm tot}(t_{\rm i})-\overline{\rm U}\right]^2, \\
\sigma_{\rm P}^2 & = & \frac{1}{{\rm I}_{\rm max}}\,\sum_{1}^{{\rm I}_{\rm max}}\, \left[{\rm P}_{\rm tot}(t_{\rm i})-\overline{\rm P}\right]^2.
\end{eqnarray}

\noindent Additionally, the light-curves are commonly measured in terms of magnitudes, so it is convenient to introduce a magnitude variable, $\Delta {\rm m}$ (as distinct from ``$m_{\rm c}$'' for clump mass), related to the model fluxes and their variations.  Since the average of the flux variations is nonzero, we defined the zero point for the light curve in magnitudes $\Delta {\rm m} = 0$ as corresponding to the relative flux level $(f_\ast+\overline{f})/f_\ast = 1+\overline{f}/f_\ast$.  Then the light curve in differential magnitude is

\begin{equation}
\Delta {\rm m}(t) = -2500\,\log\left[\frac{f_{\rm tot}(t)}{f_\ast+\overline{f}}\right],
\end{equation}

\noindent where the factor of 2500 anticipates, based on the {\em BRITE} study of WR\,40, that differential magnitude will be in milli-magnitudes (mmags).  Since $\overline{\Delta {\rm m}}=0$ by definition, the standard deviation of the photometric light curve becomes

\begin{equation}
\sigma^2_{\rm m} = \frac{1}{{\rm I}_{\rm max}}\,\sum_{1}^{{\rm I}_{\rm max}}\, \Delta {\rm m}^2_{\rm i} .
\end{equation}

Finally, it is traditional to use the polarised flux relative to the total light flux to define the polarisation.  Our assumption is that the total light is approximately equal to the direct starlight. Equation~\ref{eq:fp} implies that P = $f_{\rm P}/f_\ast$. 
Similarly, we introduce the scale constant P$_L = f_L/f_\ast$.

\subsection{Simulated Light-Curves}

Using the model described in the preceding section, we calculated a suite of simulated light-curves for photometry and polarisation.  The synthetic data were evaluated at the sampling of the observations, in terms of both sample size (number of observations) and cadence (temporal sampling).  From a given simulation of $\Delta {\rm m}(t)$ and P$(t)$, we calculate an average polarisation $\overline{\rm P}$, standard deviation in the polarisation $\sigma_{\rm P}$, 
and standard deviation in the photometry $\sigma_{\rm m}$.  

The assumed stochastic nature of the clumped wind means that we cannot evaluate the model by its ability to match each observed data point;  instead, we seek to replicate the statistical properties of the data set.  This requires us to assess the spread in the statistical properties of the simulations.  

Recalling from \cite{Ram19} that Case A models involved a distribution of clump masses, and Case B models involved an ensemble of equal clump masses, we found that both cases gave similar statistical properties for photometric and polarimetric light curves.  The addition of polarimetric data does not appear to break this degeneracy reported in the BRITE study.  While not to imply that clumps do not obey a power-law distribution,
we have chosen for simplicity to limit discussion to Case B models of equal-mass clumps in what follows.

We adopt the solution of \cite{Ram19} in terms of clump injection rate $\dot{\cal N}$ and $f_0$ for the total light variations.  For polarimetry the power-law exponent $\alpha$ appears to be a free parameter.  However, as seen in Figure~\ref{fig:mstats} from the Appendix, the key diagnostic ratio of $\overline{\rm P}/\sigma_{\rm P}$ is largely constant for a wide range in $\alpha$.  To constrain the exponent, we adopt a standard $\beta=1$ wind velocity law to infer $\alpha \approx 0.06$ (also in the Appendix).  

\begin{figure}
	\vskip -1.2truecm
	\includegraphics[width=\columnwidth]{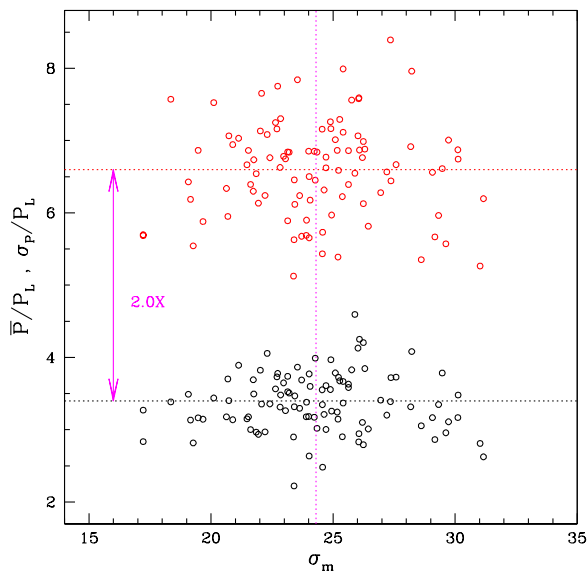}
	\vskip -2truecm
    \caption{The results of 100 model runs with  exponent value
    of $\alpha = 0.06$ for a wind consisting of identical clumps.  The horizontal is the standard deviation in the photometry of the synthetic light curves, $\sigma_{\rm m}$ in mmags.  The vertical is for the average polarisation: $\overline{\rm P}$ of a light curve relative to the scale constant, P$_L$, as the red points; the black points are for the standard deviation of the polarised light curves, $\sigma_{\rm P}$, also relative to the scale constant.  The lines signify the centroid locations for the two distributions of model runs.}
    \label{fig:m06}
\end{figure}

Adopting this value, Figure~\ref{fig:m06} displays the statistical characteristics for 100 simulated light curves.
The horizontal axis is $\sigma_{\rm m}$.  The vertical magenta line represents the observed value from the {\em BRITE} data, and assumes the same time constant, $\tau$, from the study of \cite{Ram19}.  Red points are for $\overline{\rm P}$  and black points are for $\sigma_{\rm P}$.  The model assumes thin electron scattering for the polarisation, which involves a scaling constant P$_L$ that is a free parameter of the model.  The data reveal that $\overline{\rm P}$ is about twice as large as $\sigma_{\rm P}$.  We require that the observed $\sigma_{\rm m}$ is the average of the 100 model runs.  The observations show that $\overline{\rm P} /  \sigma_{\rm P} \approx 2.1$.  Our models typically produce a ratio of around 2 for a wide range of values in $\alpha$; for Figure~\ref{fig:m06} with $\alpha=0.06$, the ratio is 2.0 as shown.

\begin{figure}
	\vskip -1.2truecm
	\includegraphics[width=\columnwidth]{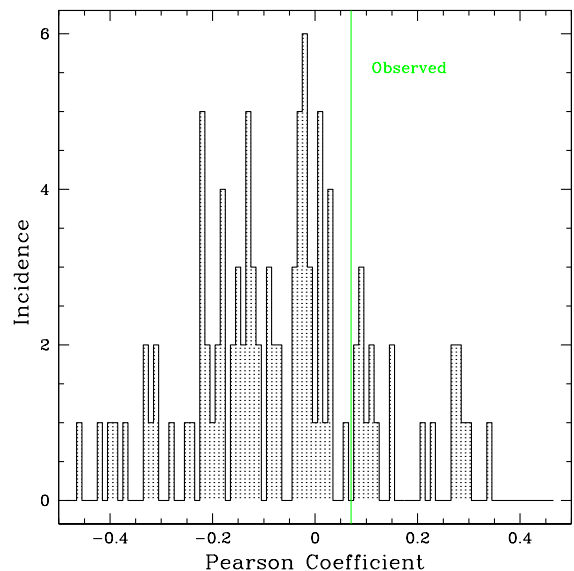}
	\vskip -2truecm
    \caption{Histogram of Pearson correlation coefficients, $r_{\rm Pearson}$, computed for $\Delta {\rm m}$ versus P for the 100 models used in Figure~\ref{fig:m06}. The vertical green 
    line is the observed value.}
    \label{fig:correlation}
\end{figure}

Another constraint provided by the data is the relative lack of correlation between polarimetric and photometric variations.  Using the Pearson correlation coefficient, Figure~\ref{fig:correlation} shows a histogram for values of $r_{\rm Pearson}$ calculated for the models in Figure~\ref{fig:m06}, as labelled.  The vertical green line is the observed value of $r_{\rm Pearson} = 0.07$ from the observations (cf., Fig.~\ref{fig:Pmag}). 

\begin{figure}
    \vskip -1.2truecm
	\includegraphics[width=\columnwidth]{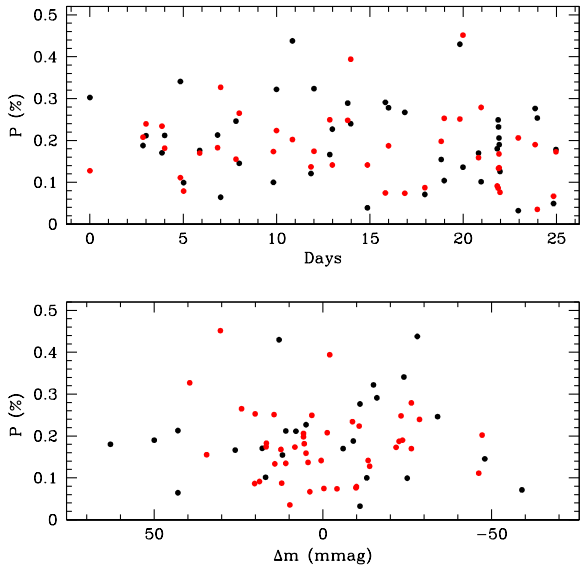}
	\vskip -2.0truecm
    \caption{A random model from the 100 runs with statistical properties shown in Figure~\ref{fig:m06}. The upper panel shows the model polarised light curve (red points) at the time sampling of the data (black points).  This comparison uses P$_L=0.03\%$ so that $\overline{\rm P}$ of the model matches that of the light curve.  The lower panel displays a plot of both observed and simulated light-curve points in polarisation plotted against the photometric variations.}
    \label{fig:particular}
\end{figure}

While Figures~\ref{fig:m06} and \ref{fig:correlation} characterise the statistical properties of the model in relation to our  data, Figure~\ref{fig:particular} shows a direct comparison between the data and a randomly selected simulated light curve from the 100 runs in Figure~\ref{fig:m06}.  The upper panel compares the observed polarisation (black) against the simulation (red).  The lower panel is a plot of the polarisation versus the photometry, again with black for the data and red for the model.  The average observed polarisation is $\overline{\rm P} = 0.199\%$; the model average is $\overline{\rm P} /{\rm P}_L = 6.6$, so P$_L = 0.03\%$ was used to scale the model polarisation light curve to match the mean polarisation of the data.

\begin{figure}
	\includegraphics[width=.9\columnwidth]{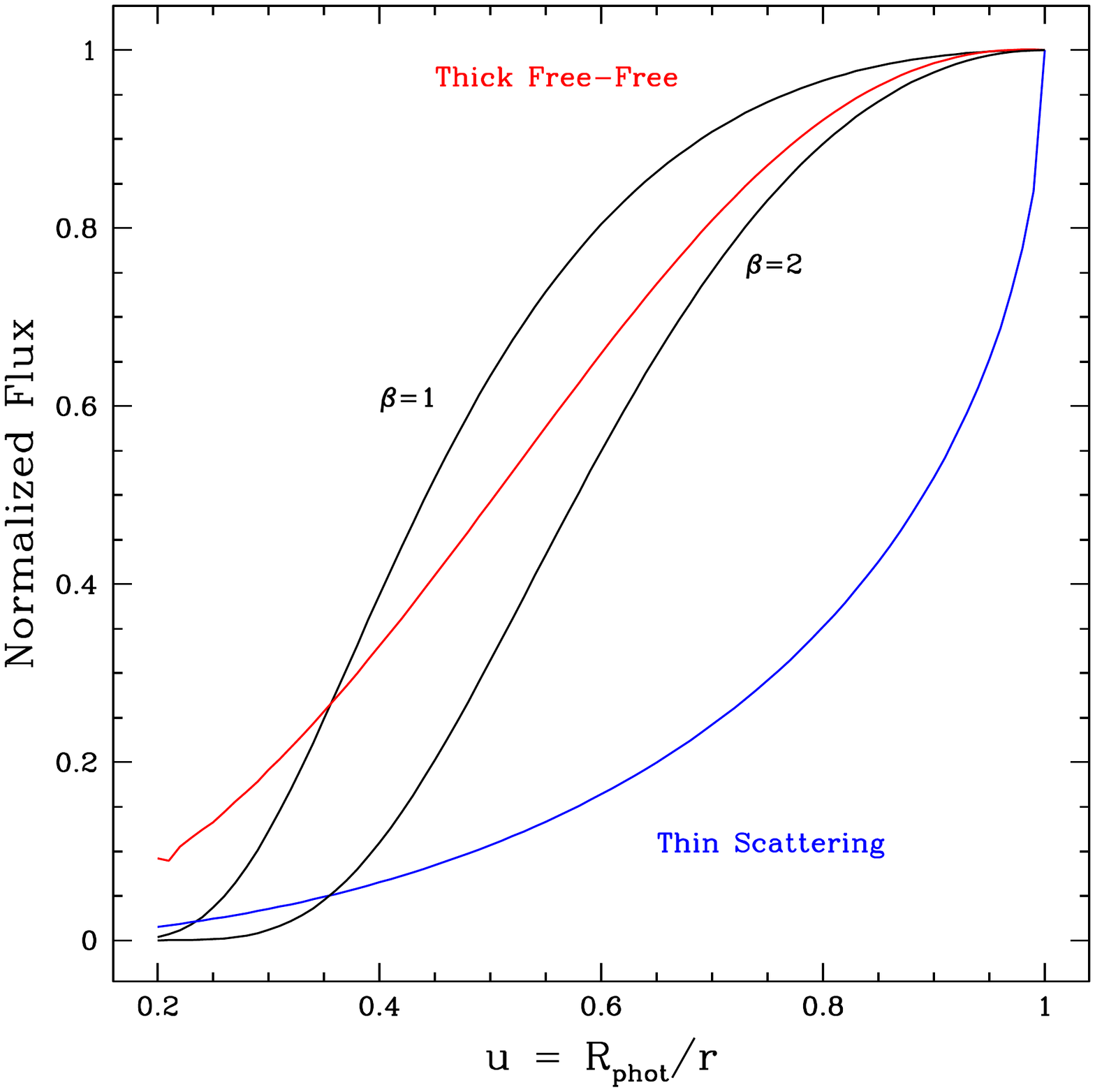}
	\includegraphics[width=.9\columnwidth]{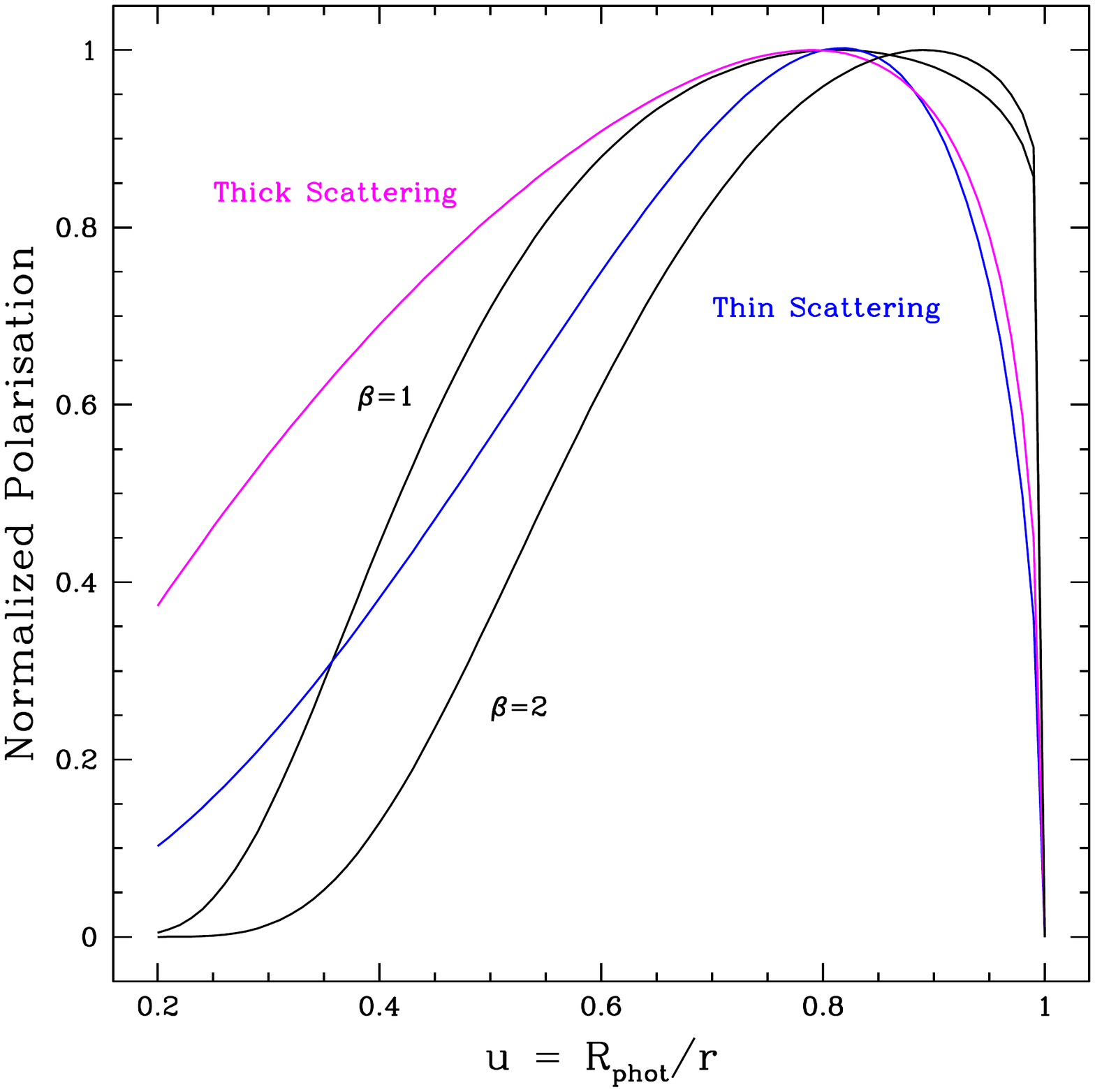}
	\vskip -1.truecm
    \caption{Comparisons between the adopted functions for how individual clumps varying in brightness mapped into the radius coordinate, here displayed as inverse radius ($u=R_{\rm phot}/r$).
    Top is for flux normalised to peak, and bottom is for polarisation normalised to peak.  At top the blue curve is for a clump that is optically thin to electron scattering, and the red curve is for a clump optically thick in free-free opacity.  The two black curves are mappings using the time profile of Equation~\ref{eq:fs} with  $\beta=1$ and $\beta=2$ velocity laws.  The panel at bottom is similar, except there is no red curve (since free-free does not produce scattering polarisation); instead, the magenta curve is for a clump that is optically thick in electron scattering.}
    \label{fig:compare2}
\end{figure}

\section{Discussion}    \label{sec:disc}

To our knowledge, previous observations of stochastic hot-star wind variability have always treated total light and polarimetry separately. Here for the first time we present nearly simultaneous data for these two modes, at least in the majority of the data. This has the advantage of eliminating epoch-dependent differential variability when comparing the two obtained at different epochs, and being able to examine the correlation, if any, between them.  One might have expected some level of anti-correlation between the light and polarisation variability.  For example, $\Delta$I might be largest and $\Delta$P smallest for clumps coming straight out at us.  The reverse is also true, that $\Delta$I would be smallest and $\Delta$P largest for a clump moving orthogonal to our line of sight. However, no correlation is observed.

Our approach has been to employ functions of time for the flux amplitude and polarisation amplitude of individuals clumps.  Additionally for polarisation, the trajectory of a clump in $\theta$ and $\phi$ is used for formulating Stokes-Q and U parameters.  Simulated light curves are obtained by randomly injecting clumps into the wind at the optical pseudo-photosphere, at a rate that maintains the mass-loss rate reported by \cite{Ham19}.



In Appendix~\ref{sec:app} we show that with selection of a velocity law for the blobs, the time functions for blob evolution can be mapped to radial coordinate.  Figure~\ref{fig:compare2} gives an example.  The upper panel is for the flux, normalised to the peak value.  The lower panel is the polarisation, also normalised to peak value.  In both cases curves for the variation in brightness or polarisation for an individual clump are plotted against inverse radius in the wind, $u = R_{\rm phot}/r$. 

Consider first the top panel.  The blue curve is for an explicitly thin scattering clump, based on Equations~8 and 10 of \cite{Ric96} with $\theta=90^\circ$.  The red curve represents a spherical clump that is optically thick to free-free opacity along its diameter, using Equation~4 of \cite{2004ApJ...610..351I}.  For this example an initial optical depth of 2.5 along the diameter was used.  Note that in order to maintain a spherical shape and an inverse square density, the radius of the clump, $R_{\rm c}$, is assumed to scale with distance in the wind according to $R_{\rm c}^3 \propto r^2$ \citep[e.g.,][]{2016MNRAS.457.4123I}.  The two black curves are mappings of Equation~\ref{eq:fs} into inverse radius using a standard $\beta$ wind velocity law.  One curve is for $\beta=1$ and the other is for $\beta=2$, as labeled. While neither of the black curves exactly matches the example of a clump optically thick to free-free opacity, both are clearly contrary to the case of a thin scattering clump which displays a precipitous drop in brightness quite near the photosphere.

Shifting to the lower panel of Figure~\ref{fig:compare2} for
polarisation, a similar set of comparisons are made.  Again, the
blue curve is for polarisation from optically thin electron scattering
using Equations~7 and 9 from \cite{Ric96}, also with $\theta=90^\circ$
as for the blue curve in the upper panel.  The curve in magenta is
for a clump that is optically thick in electron scattering and is
based on the work of \cite{1995ApJ...441..400C}.  We make use of
Equation~10 and aspects of Equation~11 from that paper.  The
polarisation for an individual clump expressed by Equation~11 is
normalised by the scattered light from the clump.  We adjust that
for normalisation by the stellar flux.  Additionally, the formulation
of those authors does not reduce to polarisation that is linear in
electron density when the clump becomes optically thin, so we have
included a bridging function to span the thin and thick scattering
limits.  With these caveats in mind, the magenta curve is for an
electron scattering optical depth of initially 2.5 across the
diameter of a spherical blob.  Again, the two black curves are for
the time function of Equation~\ref{eq:fp}, with $\alpha = 0.06$,
and $\beta=1$ and 2.  For the polarisation there is some similarity
between the black curves and the case of a thin scattering blob,
whereas there is less agreement with a blob that is optically thick
to electron scattering, where the magenta curve appears to decline
much more gradually than the black curves.

The overall impression from the comparisons in Figure~\ref{fig:compare2}
is that the total light variation of the clump implied by our model
does not seem consistent with optically thin electron scattering.
A clump that is optically thick to free-free opacity seems
characteristically more consistent with our empirical time function
mapped into the radial coordinate for the velocity laws considered.
On the other hand, our time function for polarisation is not
inconsistent with thin scattering.

Finally some considerations of clumps being all identical or obeying
a power-law distribution.  Theoretically, various studies suggest
that radially and laterally structured flow is to be expected in
massive-star winds \citep[e.g.,][]{2002A&A...383.1113D,
2003A&A...406L...1D, 2022MNRAS.tmp.2989F, 2022A&A...665A..42M}.
Empirically, Moffat \& Robert (1994) found evidence for a power law
distribution of bumps on spectral lines from WR-star winds, each
bump assumed to be from one clump (or perhaps a combination of
clumps due to so-called "nesting" effects in turbulent jargon). The
exponent $\gamma$ of the power law $N(m_c) \simeq m_c^{-\gamma}$
of 1.5 is close to what one observes in Giant Molecular Clouds,
where compressible turbulence is known to apply.  Our original hope
was that we would see a difference in our simulations of constant
clump mass {\it vs} such a power law.  The indifference of the
model to the distribution of clump properties suggests our results
in this investigation do not depend on such details, but neither
can the model provide useful constraints on this aspect of the
statistical nature of clumps.

While our modeling does not appear to distinguish between an
ensemble of identical clumps versus a power-law distribution of
masses, matching the observations does require a different number
of clumps overall for the simulation.  Modeling the BRITE light
curve for WR~40 required about $4\times$ as many identical clumps
as compared to the model with a power-law in clump masses \citep[c.f.,
Tab.~4 of][]{Ram19}.
This suggests that statistical spatial clustering of identical
clumps could perhaps approximate the approach with a diversity of
clump properties.  The study of \cite{2000A&A...356..619B} showed
that the detailed structure of wind inhomogeneities influences
polarization only when of sufficient radial and/or angular\footnote{Angular
extent referring to the angle of the wind structure from the
perspective of the stellar source of illumination.} extent.
Consequently, clustering of small clumps is expected to act
polarimetrically as a single larger clump.  

\section{Conclusions}   \label{sec:conc}

In previous attempts to explain variable brightness and/or polarimetry
from WR winds, the approach has generally been to adopt functions
of radius motivated by the expected physics.  For example regarding
the polarisation modelling, some authors adopted optically thin
electron scattering for its semi-analytic nature as a convenience
for exploring a broad range of wind and clump parameter space
\citep[e.g.,][]{Bro95, Ric96, 2000A&A...357..233L}, while others
considered numerically intensive calculations of multiple scattering
for a range of clump optical depths but limited to sparse sampling
of parameter space \citep[e.g.,][]{1995ApJ...441..400C,
2000ApJ...540..412R, 2012AIPC.1429..278T}.
However, while these methods have produced some successes, none
have demonstrated success in reproducing simultaneously the properties
of the stochastic variability expressed in terms of $\sigma_{\rm
m}$, $\overline{\rm P}$, $\sigma_{\rm P}$, $\tau$, and the apparent
absence of correlation between photometric and polarimetric variations.

The study of this paper has focused on interpreting on variable
photopolarimetry of the WN8 star WR~40.  The WN8 stars are among
the coolest WRs with the slowest winds and fit well into the general
trend of variability level with WR subtype, whether in spectroscopy
\citep{2020ApJ...903..113C}, photometry \citep{2022ApJ...925...79L},
or (broad-band) polarimetry \citep{Rob89}.  The currently best
explanation of this trend is subsurface convection, which is deeper,
hence denser and more energetic, in cooler WR stars \citep{Mic14},
which also have the slowest winds.  Among WR~stars, WR~40 was
selected specifically because of a unique dataset that displays
simultaneous photometric and polarimetric variability.

In our interpretative modeling, we chose to extend the work of
\cite{Ram19} in adopting a temporal function for clump brightness
to include variation in polarisation as well.  And we find that the
observations can, characteristically, be reproduced with such a
model, at least in the case of the WN8h star WR\,40.  The disadvantage
of the approach is the apparent detachment from underlying physics
governing the clump properties, such as shape or mass.  In principle,
the two approaches can be bridged, as illustrated by
Figure~\ref{fig:compare2}.  That particular illustration involved
the use of a wind velocity law. However, such an approach makes the
tacit assumption that a blob retains its character as it moves away
from the star.  Perhaps the clump evolves, such as fragmenting,
colliding with other clumps, or changing shape.  Given the reasonable
agreement between the measurements and the modeling approach, the
time functions that we have adopted can serve to constrain future
physical models for structured winds.

\section*{Acknowledgements}

The authors are grateful to an anonymous referee for comments that
led to improvements in the manuscript.  RI gratefully acknowledges
support from the National Science Foundation under Grant No.
AST-2009412 and NASA grants HST-GO-15822.002-A and HST-GO-16170.002-A.
AFJM, CR, and LD are grateful to NSERC (Canada) for financial
support. We thank T.~Ramiaramanantsoa for the periodicity check.

\section*{Data Availability}


The data underlying this article are available in the article and in its online supplementary material.



\bibliographystyle{mnras}
\bibliography{references-wr40} 

\appendix
\section{Constraining the Exponent 
$\alpha$} \label{sec:app}

In the study of WR\,40 by \cite{Ram19}, the free parameters of the model consisted of the flux amplitude of the clumps $f_0$, the number of clumps $N_T$, and the timescale for declining brightness of the clump $\tau$.  These parameters were constrained by the statistical properties of the variability observed in the {\em BRITE} light curve for WR\,40.  The current analysis adds two further parameters $\alpha$ and $f_L$ to model the observed variability in the linear polarisation. 

\begin{figure}
   \vskip -1.2truecm
	\includegraphics[width=\columnwidth]{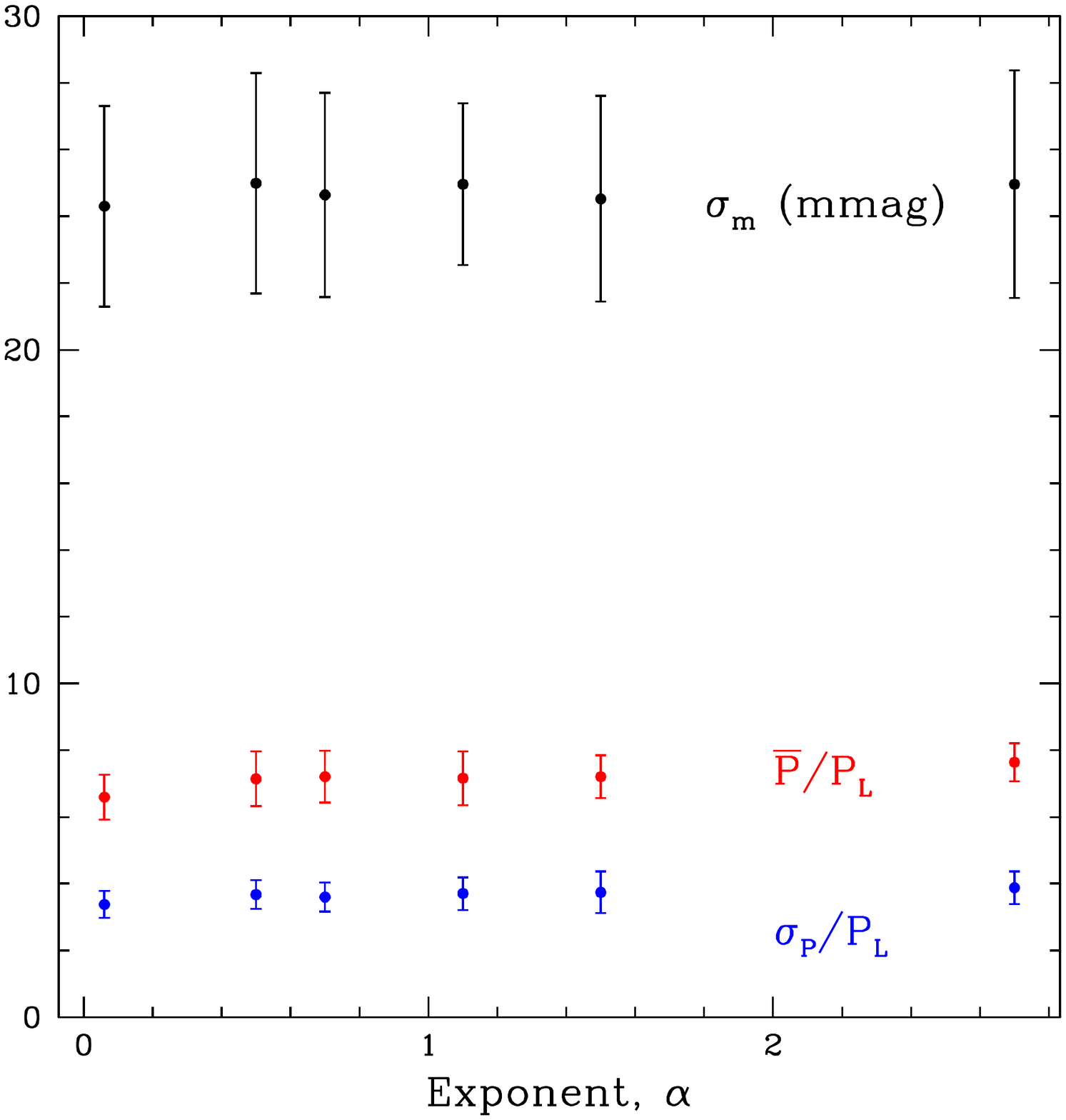}
	\vskip -2.0truecm
    \caption{A plot of the statistical properties of simulated light curves with the exponent $\alpha$.  The vertical bars represent the standard deviation for 25 model runs at each of the values of $\alpha$ used.  Black is for $\sigma_{\rm m}$ which has no functional dependence on $\alpha$.  Red is for the average polarisation $\overline{\rm P} /{\rm P}_L$.  Blue is for $\sigma_{\rm P}/{\rm P}_L$.  As can be seen, the polarimetric properties are relatively constant below $\alpha =1$, until $\alpha$ is quite small; for $\alpha >1$ the red and blue points decline.  However, the ratio $\overline{\rm P} / \sigma_{\rm P} \simeq 2$ for the values of $\alpha$ shown.}
    \label{fig:mstats}
\end{figure}

Figure~\ref{fig:mstats} shows model results for several values of the exponent $\alpha$.  The model produces both flux and polarisation light curves at the sampling of the observations.  From these the standard deviation in the photometric light curve, $\sigma_{\rm m}$, the average polarisation, $\overline{\rm P}$, and the standard deviation of the polarisation light curve, $\sigma_{\rm P}$ are found and plotted here as black, red, and blue, respectively.  The points are the averages for 25 model runs with fixed model parameters, and the error bars represent the standard deviation in those averages.  

Of course, the flux variations do not depend on the exponent $\alpha$; only polarisation depends on $\alpha$.  The observations reveal that $\overline{\rm P}/\sigma_{\rm P} \approx 2$.  This ratio is approximately achieved for the range of $\alpha$ as shown.  Moreover, $\overline{\rm P}$ and $\sigma_{\rm P}$ are constant nearly to $\alpha \simeq 1$, declining with larger values of $\alpha$.  Thus the observed ratio of $\overline{\rm P}/\sigma_{\rm P}$ does not constrain the value of $\alpha$ for the model.  The values of $\overline{\rm P}$ and $\sigma_{\rm P}$ both involve the scaling constant ${\rm P}_L$, which is determined by the data but has no impact on $\alpha$.

A constraint can be placed on $\alpha$ by the wind velocity.  The model for the variation of photometric light from a clump declines monotonically with time; however, the polarised flux starts at zero, rises to a peak, and declines thereafter.  This behaviour is reflective of the finite star depolarisation factor of \cite{Cas87}.  Considering the scattered polarised light from a clump as a function of radius, this too starts at zero, peaks, then declines.  The bridge between these two approaches is the wind velocity which determines the time of flight of a clump from the pseudo-photosphere to the radius at which peak polarisation occurs, and thus the time also.

Using Equation~\ref{eq:fp}, setting $df_{\rm P}(t)/dt =0$ gives the time of peak polarisation as

\begin{equation}
    t_{\rm peak} = \sqrt{\frac{\alpha}{2}}\,\tau .
\end{equation}

\noindent The radius-dependent formulation for the polarisation involves

\begin{equation}
    f_{\rm P}(r) \propto D(r) \, r^{-2},
    \label{eq:fpr}
\end{equation}

\noindent where $D = \sqrt{1-R_{\rm phot}^2/r^2}$ is the finite
star depolarisation factor.  Peak polarisation is achieved at the location $r_{\rm peak}/R_{\rm phot} = \sqrt{3/2} = 1.22$.

Relating the peak in time with the peak in radius is achieved via

\begin{equation}
    t_{\rm peak} = \int_{R_{\rm phot}}^{r_{\rm peak}} \, \frac{dr}{v(r)},
\end{equation}

\noindent where $v(r)$ is the wind velocity law.  For this we use a standard $\beta=1$ velocity law with

\begin{equation}
    v(r) = v_\infty\,\left( 1 - b\, R_\ast/r \right)
        = v_\infty\,\left( 1 - b'\, R_{\rm phot}/r \right) ,
\end{equation}

\noindent where the second expression renormalises the radius $r$ in terms of the photospheric radius, introducing $b' = bR_\ast/R_{\rm phot}$.  From $R_{\rm phot} =2.4 R_\ast$ and $b \approx 0.9$, we find  $b' \approx 0.4$. 

Setting the two expressions for $t_{\rm peak}$ equal and introducing $t_\infty = R_\ast/v_\infty$ yields

\begin{equation}
    \alpha \approx 1.4\, (t_\infty/\tau)^2 \approx 0.06.
\end{equation}

\bsp	
\label{lastpage}
\end{document}